# Optical Nanotransmission Lines: Synthesis of Planar Left-Handed Metamaterials in the Infrared and Visible Regimes

Andrea Alù

University of Pennsylvania, Dept. of Electrical and Systems Engineering, Philadelphia, PA 19104, U.S.A.
University of Roma Tre, Dept. of Applied Electronics, Rome, Italy

Nader Engheta

University of Pennsylvania, Dept. of Electrical and Systems Engineering, Philadelphia, PA 19104, U.S.A.

**Abstract**

Following our recent theoretical development of the concept of nano-inductors, nano-capacitors and nano-resistors at optical frequencies and the possibility of synthesizing more complex nano-scale circuits, here we theoretically investigate in detail the problem of optical nano-transmission-lines (NTL) that can be envisioned by properly joining together arrays of these basic nano-scale circuit elements. We show how, in the limit in which these basic circuit elements are closely packed together, the NTLs can be regarded as stacks of plasmonic and non-plasmonic planar slabs, which may be designed to effectively exhibit the properties of planar metamaterials with forward (right-handed) or backward (left-handed) operation. With the proper design, negative refraction and left-handed propagation are shown to be possible in these planar plasmonic guided-wave structures, providing possibilities for sub-wavelength focusing and imaging in planar optics, and laterally-confined waveguiding at IR and visible frequencies. The effective material parameters for such NTLs are derived, and the connection and analogy between these optical NTLs and the double-negative and double-positive metamaterials are also explored. Physical insights and justification for the results are also presented.

**Introduction**

Left-handed metamaterials (also known as double-negative (DNG) media, in which the real parts of permittivity and permeability are both negative in a given range of frequencies) and single-negative (SNG) materials (with the real part of the permittivity negative ($\varepsilon$-negative, ENG) or of the permeability negative ($\mu$-negative, MNG)) offer exciting electromagnetic wave propagation properties, with potential applications in the design of various devices and components (see e.g., [1]-[7]). The resonance phenomenon present at the interface between a DNG material and a conventional material (which in analogy with the previous terminology we name "double positive", DPS), and/or between ENG and MNG media [1], may be manifested in several interesting ways, e.g., in sub-wavelength focusing and 'perfect' lensing [2]. In our recent work, we have also shown how the suitable pairing of DNG and DPS materials may overcome the diffraction limit in different configurations such as in closed and open waveguides and cavities [4]-[6], or in scattering and radiation from electrically small objects [7]-[8], to name a few. The realization of DNG metamaterials, although achieved by several groups (see e.g., [9]), requires technological efforts and challenges at the present time. At optical frequencies, moreover, the possibility of inducing a negative effective permeability in a bulk material is even more challenging [10] and different alternatives have been proposed in the recent literature for constructing left-handed metamaterials in the optical regime [11]-[16]. This is why in some of our previous work (e.g., [1], [4]-[7]), we have investigated single-negative (SNG) media, such as ENG plasmonic material (e.g., noble metals at infrared (IR) and optical frequencies, or ENG artificially engineered metamaterials at lower frequencies), some of whose properties are somewhat similar to those of DNG media. Also the technique we have proposed to realize an effective µ-negative material in the optical regime relies on properly exploited plasmonic resonances of ENG nanoparticles [16]. Moreover, open ENG planar layers or cylinders supporting guided surface modes may overcome the diffraction limit in their operation [6], [17]-[18], i.e., the more the waveguide's lateral transverse cross section is reduced, the more the fundamental mode remains concentrated around the waveguide, implying in principle the existence of a mode with lateral cross section below the conventional $\lambda/2$ limit, which may hint to the possibility of miniaturization of optical interconnects.

We have also recently shown theoretically that nano-scale circuit elements such as nano-inductors and nano-capacitors at optical frequencies can be conceptualized using properly designed plasmonic (ENG) and DPS nano-particles small compared with the operating wavelength [19]. Since their dimensions are assumed to be electrically small, for the spatial distribution of the fields in their vicinities their interaction with light may indeed be treated as "quasi-static" and it may be interpreted as "lumped circuit elements" at these frequencies. The role of the conduction current in a conventional circuit is here replaced by the displacement current circulating around and inside the nano-particles. In that work, we have also speculated the possibility of designing more complex optical nano-scale circuits by combining together these basic elements in *series* and *parallel* configurations. We are currently investigating those issues.

Our proposed use of plasmonic and non-plasmonic nanoparticles as nano-scale circuit elements naturally leads to the possibility of synthesizing a nano transmission line (NTL) at the optical frequencies when one considers combining together the basic nano-inductors and nano-capacitors proposed in [19], as we have anticipated in a recent conference presentation [20]. As we will show in the following, this may also provide interesting possibilities for the realization of one- and two-dimensional planar negative-refraction metamaterials at optical frequencies. In the present work we will first briefly review the concept of nano-scale optical circuit elements in order to introduce the possibility of synthesizing a right-handed or left-handed NTL by placing in close proximity DPS and plasmonic nano-particles. Then, we will show how in the limit in which these elements are closely packed together, the NTL geometry may evolve into the planar open waveguide we have recently analyzed [6]. The analysis of this structure will be presented, showing how the supported modes may be either forward-wave or backward-wave modes, and therefore how these planar layers may synthesize an effective right-handed (RH) or left-handed (LH) 2-D planar metamaterial at optical frequencies. It is possible therefore to predict negative refraction at the interface between such RH and LH 2-D NTL and consequently sub-wavelength focusing due to such negative refraction. Extending this concept, we also foresee the possibility of synthesizing SNG materials with the desired properties in similar geometries, thus potentially obtaining other anomalous properties of the wave interaction with these materials. The SNG version of these structures is currently under study and will be reported in a future publication

## Optical Nano-Transmission Lines as Collections of Nano-Inductors and Nano-Capacitors

It is well known that with a $e^{-i\omega t}$ monochromatic excitation the complex impedance of a lumped circuit element represents the ratio between the voltage phasor applied at its two terminals and the conducting current flowing through the element. In the range of amplitudes for which the element behaves linearly, these two quantities are proportional and the impedance is easily defined and measured as the proportionality constant. For lossless "reactive" elements this quantity is purely imaginary and provides the information for the phase advance or delay of the voltage with respect to the current for the inductive or capacitive reactance. Of course, lossy and absorbing passive elements have a positive real part for the impedance, taking into account the dissipated power. At optical frequencies, however, "usual" lumped inductors and capacitors are not available, since the conductivity of materials (particularly metals) is modified in these frequency regimes. In [19], however, we have theoretically proposed how a single nano-particle, small compared to the operating wavelength, may behave as an equivalent "lumped" circuit element, as seen from outside, provided we consider the electric *displacement* current density ($-i\omega\varepsilon E_0$, with $\varepsilon$ being the local electric permittivity and $E_0$ the local electric field) as the circulating current in this nano-element and in the space surrounding it. In the quasi-static limit, in fact, the "average" voltage across the element and the total displacement current induced all over the space satisfy Kirchhoff's laws and the effective impedance of a spherical nano-particle of radius $a$ may be defined as [19]:

$$Z = \left(-i\omega\varepsilon\pi a\right)^{-1}. \tag{1}$$

Similar formulas may be derived for more complex shapes for the nano-particles. In particular, a DPS nano-particle corresponds to a nano-capacitor, since its permittivity is positive, whereas a plasmonic nano-particle acts as a nano-inductor as observed from outside, due to its negative permittivity. We also speculated that two closely packed nano-particles, by the same token, may be regarded as in *parallel* (having the same voltage applied) or in *series* (having the same current) configuration, depending on the orientation and polarization of the exciting field. It may be straightforward then to suggest more complex circuits when a system of nano-particles is properly arranged [19]. The circuit model of a "right-handed (RH)" transmission-line is well known in the electrical engineering community, and as depicted in Fig. 1 (top row, left column), consists of the cascade of distributed series inductors and shunt capacitors. This schematic circuit supports a transverse electromagnetic (TEM) wave with RH (i.e., forward-wave) propagation and it represents a useful circuit model for commonly used guiding structures at microwave and lower frequencies (e.g., coaxial cables and striplines). Interchanging the role of inductors and capacitors, one may synthesize a left-handed (LH) transmission-line, as depicted in Fig. 1 (top row, right column). This circuit supports LH (backward-wave) propagation and this model has been exploited by several groups to synthesize 2-D planar non-resonant LH metamaterials at microwave frequencies [21]-[22] with backward-wave propagation to verify negative refraction and sub-wavelength focusing at these frequencies. In the figure (first row) are also depicted arrows to show how voltages ($V$) and currents ($I$) along the line are defined. Can one have a similar structure at the IR and optical frequencies? If one can consider nano-inductors and nano-capacitors, an analogous nano-transmission line (NTL) can be synthesized at these frequencies, as we envision here.

Following the circuit analogy described above, it may be possible to arrange properly the nano-particles acting as nano-inductors and nano-capacitors in order to synthesize a NTL at optical frequencies, as sketched in Fig. 1 (middle row). In the figure, darker spheres represent DPS nano-particles acting as nano-capacitors, whereas the lighter ENG particles play the role of nano-inductors. Also in this case an analogous representation for voltages and currents is suggested, even though here they symbolize different, but related, physical quantities. In the limit in which such elements are closely packed together, such configurations may eventually resemble stacks of planar layers of DPS and plasmonic materials (bottom row in Fig. 1), which correspond to the geometry we have studied in [6] to show the anomalies in the surface-wave propagation when DNG materials are involved. Motivated by these analogies, in the next sections we discuss the modal distribution of a stack of planar layers of ENG and DPS materials, showing how it is indeed possible to synthesize positive-index and negative-index sub-diffraction optical NTL by using such plasmonic open planar waveguides, and how such 2-D positive-index/negative-index structures may be employed in a similar way as their double-positive (DPS) and double-negative (DNG) metamaterial counterparts in order to achieve negative refraction and sub-wavelength focusing. This may open doors to interesting potential in imaging, lensing and waveguiding applications at optical and infrared, as well as microwave, frequencies.

## 2-D Planar Nano Transmission Lines

The structures in the last row of Fig. 1 may be envisioned as coaxial cylindrical rods or stacked planar slabs of DPS and ENG materials, giving rise to 1-D or 2-D NTL, respectively. In the following, we analyze in detail the planar slab configuration, which may be utilized to synthesize right-handed (RH) and left-handed (LH) 2-D NTL, in analogy with what has been obtained at microwave frequencies by lumped circuit elements for construction of 2-D planar metamaterials [21]-[22]. It is also important to note that a similar configuration, although obtained following a different train of thoughts, has been envisioned by Shvets in [15] as a first step for constructing a photonic band-gap material with negative-refractive properties in optics. We will see later, however, how its operation is different in several aspects from what we present here.

The geometry we refer to is depicted in Fig. 2 and for simplicity it consists of a planar slab with permittivity $\varepsilon_{in}(\omega)$ and thickness $d$ sandwiched between two half-spaces with permittivity $\varepsilon_{out}(\omega)$, at operating frequency $f = \omega/(2\pi)$. The slabs are infinitely extent in the *x-z* plane and the materials considered are non-magnetic i.e., their permeabilities are taken to be the free space permeability $\mu_0$. The complex permittivities $\varepsilon_{in}$ and $\varepsilon_{out}$ are in general dependent on frequency.

Consider first the guided modes of this structure propagating along the $x$ direction with a factor $e^{i\beta x}$. Due to the symmetry of the structure, the guided modes may be categorized into even and odd modes with respect to the transverse variation along the $y$ axis and into TE and TM modes with respect to the direction of propagation. In this case, therefore, all quantities are independent of $z$ variable.

A supported TM mode has the following magnetic field distribution, depending on whether it is even or odd:

$$\mathbf{H}_{\substack{even \\ odd}} = \hat{\mathbf{z}} H_0 e^{i\beta x} \begin{cases} \sinh\left[\sqrt{\beta^2 - \omega^2 \varepsilon_{out} \mu_0}\, d/2\right] e^{-\sqrt{\beta^2 - \omega^2 \varepsilon_{out} \mu_0}\,(y-d/2)} & y > d/2 \\ \cosh\left[y\sqrt{\beta^2 - \omega^2 \varepsilon_{in} \mu_0}\right] & |y| < d/2 \\ \sinh\left[y\sqrt{\beta^2 - \omega^2 \varepsilon_{in} \mu_0}\right] & \\ \pm \sinh\left[\sqrt{\beta^2 - \omega^2 \varepsilon_{out} \mu_0}\, d/2\right] e^{\sqrt{\beta^2 - \omega^2 \varepsilon_{out} \mu_0}\,(y+d/2)} & y < -d/2 \end{cases}, \quad (2)$$

$H_0$ is the arbitrary complex amplitude of the magnetic field. The corresponding electric field may be derived at every point from Maxwell equations: $\mathbf{E} = \dfrac{\nabla \times \mathbf{H}}{-i\omega\varepsilon}$.

Moreover, the guided wave number $\beta$ satisfies the following dispersion relations for even and odd modes:

$$even: \tanh\left[\sqrt{\beta^2 - \omega^2 \varepsilon_{in}\mu_0}\,\frac{d}{2}\right] = -\frac{\varepsilon_{in}}{\varepsilon_{out}}\frac{\sqrt{\beta^2 - \omega^2 \varepsilon_{out}\mu_0}}{\sqrt{\beta^2 - \omega^2 \varepsilon_{in}\mu_0}}$$
$$odd: \coth\left[\sqrt{\beta^2 - \omega^2 \varepsilon_{in}\mu_0}\,\frac{d}{2}\right] = -\frac{\varepsilon_{in}}{\varepsilon_{out}}\frac{\sqrt{\beta^2 - \omega^2 \varepsilon_{out}\mu_0}}{\sqrt{\beta^2 - \omega^2 \varepsilon_{in}\mu_0}}$$
(3)

The TE mode equations may be easily obtained by duality.

When the permittivity of one of the two materials has negative real part (e.g., plasmonic materials), i.e., when $\text{sgn}[\text{Re}\,\varepsilon_{out}] = -\text{sgn}[\text{Re}\,\varepsilon_{in}]$, these guided modes show peculiar properties. First of all, the fundamental TM mode does not have in principle a cut-off thickness: while decreasing the core thickness $d$ to sub-wavelength dimensions, still Eq. (3) admits a real solution for $\beta$ in the limit of no losses. As was shown in [6] for the DNG slab and also similarly to the cylindrical case [17]-[18], the wave number $\beta$ increases as $d$ decreases, causing concentration of the guided field distribution around the two interfaces $y = \pm d/2$ (as can be implied from Eq. (2) when the slab thickness reduces. (Of course, a physical/practical limit may still hold, since such a slow mode, concentrated in an electrically small cross section, would be highly sensitive to the ohmic losses present in any realistic material.) Neglecting material losses for the moment and assuming sub-wavelength thickness for the core slab, i.e., $d \ll \min\left[\frac{2\pi}{|\omega\sqrt{\varepsilon_{in}\mu_0}|}, \frac{2\pi}{|\omega\sqrt{\varepsilon_{out}\mu_0}|}\right]$, if the corresponding $\beta$ is much larger than $\max\left[|\omega\sqrt{\varepsilon_{in}\mu_0}|, |\omega\sqrt{\varepsilon_{out}\mu_0}|\right]$, Eq. (3) can be simplified into the following simple relations:

$$even: |\beta| = -\frac{2}{d}\tanh^{-1}\frac{\varepsilon_{in}}{\varepsilon_{out}}$$
$$odd: |\beta| = -\frac{2}{d}\coth^{-1}\frac{\varepsilon_{in}}{\varepsilon_{out}}$$
(4)

which allow an easy determination of the wave number dispersion for the geometry of Fig. 2. As an aside, we note from Eq. (4) that, since $d > 0$ and $\beta$ is real, it follows that in this sub-wavelength lossless case the constraint $-1 < \varepsilon_{in}/\varepsilon_{out} < 0$ for even modes and $\varepsilon_{in}/\varepsilon_{out} < -1$ for odd ones should hold.

In Fig. 3, we show the dispersion plots for four different configurations: in Fig. 3a, $\varepsilon_{in} = -5\varepsilon_0$ and $\varepsilon_{out} = \varepsilon_0$ (odd mode) and reversing the two materials (even mode), and in Fig. 3b $\varepsilon_{in} = \varepsilon_0$ and $\varepsilon_{out} = -\varepsilon_0/5$ (odd mode) and reversing the two materials (even mode). In these figures, we also compare the plots of the exact Eq. (3) and the approximate Eq. (4) valid for sub-wavelength thicknesses. The wave number in all cases increases hyperbolically with the reduction of thickness, as predicted by Eq. (4). Notice that by increasing the waveguide thickness beyond a certain threshold (which depends on the material parameters, and is beyond the limits plotted in this figure) the wave number $\beta$ for the dispersion curves related to the geometries with air in the cladding region ($\varepsilon_{out} = \varepsilon_0$) reaches the limit $\beta = \omega\sqrt{\varepsilon_0\mu_0}$, at which point the fundamental mode we are considering here is no longer guided. Other higher-order modes may be guided, however, by such a thick slab, but we will not discuss those modes here, since we are interested in the sub-wavelength mode that follows Eq.(4), and for which the previous heuristic circuit analogy (valid when the nano-particles act as nano-circuit elements) holds.

Still assuming lossless materials, for guided modes $\beta^2 - \omega^2\varepsilon_{out}\mu_0 > 0$ and the local Poynting vector along the direction of propagation ($S_x$) is a real quantity with opposite signs in the two materials. This implies that two oppositely-signed power fluxes are present inside ($P_{in}$) and outside ($P_{out}$) the core slab with the following amplitudes in the TM case:

$$P_{out}^{\substack{even \\ odd}} = \frac{\beta|H_0|^2}{4\omega\varepsilon_{out}} \frac{\cosh^2\left[\sqrt{\beta^2 - \omega^2\varepsilon_{in}\mu_0}\,\frac{d}{2}\right]}{\sinh^2\left[\sqrt{\beta^2 - \omega^2\varepsilon_{in}\mu_0}\,\frac{d}{2}\right]}\,\frac{1}{\sqrt{\beta^2 - \omega^2\varepsilon_{out}\mu_0}}$$

$$P_{in} = \frac{\beta|H_0|^2}{4\omega\varepsilon_{in}}\left(\frac{\sinh\left[\sqrt{\beta^2 - \omega^2\varepsilon_{in}\mu_0}\,d\right]}{\sqrt{\beta^2 - \omega^2\varepsilon_{in}\mu_0}} - d\right)$$

(5)

Eq. (5) ensures that for positive $\beta$, i.e., phase velocity parallel to the positive $x$ axis, $\text{sgn}\,P_{out} = \text{sgn}\,\varepsilon_{out}$ and $\text{sgn}\,P_{in} = \text{sgn}\,\varepsilon_{in}$, which shows how the two fluxes flow in opposite directions (since $\text{sgn}\,\varepsilon_{in} = -\text{sgn}\,\varepsilon_{out}$). In particular, the *forward* power flux, flowing parallel to the phase velocity, resides in the DPS region (with $\varepsilon > 0$) and the *backward* one, flowing anti-parallel to the phase velocity, in the ENG plasmonic one (with $\varepsilon < 0$), independent of the relative position of the two materials. The net-power flux evidently consists of the algebraic sum $P_{net} = P_{in} + P_{out}$, which in magnitude is less than $\max(|P_{in}|, |P_{out}|)$. Considering a source positioned somewhere along the waveguide for exciting this mode, an observer would see part of the power taken from the source in one of the two regions flowing back to the source in the other region, with no transverse exchange of power between the two fluxes in the steady-state scenario. However, as discussed in [6], [23] for the case of closed waveguides partially filled with SNG or DNG materials with similar properties, this does not present any problem for causality, since this anomalous feature is explained and justified by considering the reflection of the mode at any discontinuity in a realistic finite structure (The reader is referred to [6], [23] for a more detailed explanation of the phenomenon).

For positive $\beta$ when we get $P_{net} > 0$ the corresponding mode is a *forward* mode, since its phase velocity is parallel with the net-power flow (and thus with its group velocity); if instead $P_{net} < 0$, we are dealing with a *backward* mode, with anti-parallel phase and group velocities. The special case of $P_{net} = 0$, i.e., $P_{in} = P_{out}$ and zero group velocity, would correspond to an anomalous resonating mode with non-zero phase velocity, similar to what we have already found in [5]-[6], [23]. As we will show later, however, this case is possible here only in the limit of $\varepsilon_{in} = -\varepsilon_{out}$, which leads to a plasmonic surface resonance at each of the two interfaces $y = \pm d/2$.

It is interesting to note that the forward or backward behavior of the supported mode may be predicted directly from Eq. (4) considering the group velocity $v_g = \partial \omega / \partial \beta$. Considering waveguides in which the plasmonic material has permittivity $\varepsilon_{ENG}(\omega)$ and the DPS material is vacuum with permittivity $\varepsilon_0$ (this may also be another DPS material with non-dispersive behavior), for positive phase velocity ($\beta > 0$) when we take the derivative of (4) with respect to the frequency, we get:

$$\frac{\partial \beta}{\partial \omega} = \frac{1}{v_g} = \frac{2\varepsilon_0}{d\left(\varepsilon_{ENG}^2 - \varepsilon_0^2\right)} \frac{d\varepsilon_{ENG}}{d\omega}, \tag{6}$$

which is valid for both even and odd modes, and for air-ENG-air and ENG-air-ENG waveguides. Remembering that in not-highly-absorptive regions $d\varepsilon / d\omega$ is a strictly positive quantity in any passive material (see [10], §84), it follows that a positive group velocity, and therefore a forward mode, is obtained when $\varepsilon_{ENG} < -\varepsilon_0$ and a backward mode is supported for $-\varepsilon_0 < \varepsilon_{ENG} < 0$. This is justified by the fact that when $|\varepsilon_{ENG}| > \varepsilon_0$ the mode is less distributed in the ENG material and more "available" in the DPS one, independent of their relative position. As a result, the negative power flow present in the ENG material is also less than the positive one in the DPS and the net power remains positive, i.e., parallel to the phase velocity. For backward modes the situation is reversed, since $|\varepsilon_{ENG}| < \varepsilon_0$. This interesting feature confirms our heuristic NTL-motivated prediction following which we came out with this geometry. That is if we consider even modes, which are those consistent with the TL model of Fig. 1 [24], and we require $-1 < \varepsilon_{in} / \varepsilon_{out} < 0$, we will find that an ENG-DPS-ENG waveguide would have $\varepsilon_{ENG} < -\varepsilon_0$ and therefore a forward-wave behavior as predicted by Fig. 1a, whereas an DPS-ENG-DPS waveguide will have $-\varepsilon_0 < \varepsilon_{ENG} < 0$ and therefore a backward behavior, as in Fig. 1b. In these sub-wavelength structures the field distributions are very much concentrated around the interfaces $y = \pm d/2$, in some sense resembling a TL made of conducting wires running along $y = \pm d/2$ at lower frequencies. In other words, here we may envision RH or LH NTLs at optical frequencies. Applying similar consideration, the odd configuration would behave in a dual way, as evident from Eq. (6). This is consistent with what discussed by Shvets in [15] for the plasmonic-air-plasmonic waveguide supporting an odd backward mode.

Fig. 4 shows the field distribution, for a RH (i.e., positive-index) (on the left) and a LH (i.e., negative-index) (on the right) NTL, designed to have the same thickness for the core slab, which is $d = \lambda_0 / 50$. (The issue of equal thickness for the core slabs becomes important when we want to cascade and match an RH with an LH NTL, as we will discuss in the next section.) The RH NTL is formed with an ENG cladding with $\varepsilon_{out} = -1.05\varepsilon_0$ surrounding a vacuum core, whereas the LH NTL is formed with a core ENG slab with $\varepsilon_{in} = -0.95\varepsilon_0$ in empty space. The supported $\beta$ for the modes depicted in the figure are $\beta_{RH} = 29.5k_0$ and $\beta_{LH} = 28.95k_0$, respectively. You may notice how the two field distributions resemble each other (this is of course designed intentionally in order to ensure a good matching between the two NTLS when they are cascaded, as will be explained in the next section). For a positive direction of power flow, the two wave numbers $\beta_{RH}$ and $\beta_{LH}$ need to have opposite signs.

We emphasize again how the field distributions may somehow "resemble" those of a standard TL, with "currents" concentrated in the region near the two interfaces. Note, however, how the "currents" are now distributed in the space, and by decreasing the core thickness $d$ the effective region they occupy becomes proportionally narrower (the fact that $\beta$ and $d$ are inversely proportional as in Eq. (4) ensures that a scaling in the structure corresponds to an anomalous scaling in the transverse field distribution). Consistent with our circuit analogy, these "currents" may now be represented by the integral of the displacement current $-i\omega\varepsilon E_x$ around the two interfaces. They are indeed oppositely oriented for the two interfaces, as it is the case in a standard TL. (We note, however, that in this case each of the two currents actually consists of two opposite fluxes running around each of the two interfaces, similar to what we have already discussed for the power flow.) Applying Ampere's law, the integral sum of these distributed displacement currents around each interface has the value:

$$I = \int_0^\infty -i\omega\varepsilon E_x \, dy = \int_0^\infty \frac{\partial H_z}{\partial y} dy = -H_z(y=0). \qquad (7)$$

(Note that this current has the unit of A/m, since this is a current per unit length of the structure in the z direction.) The phase velocity of the modes is much slower than that of a plane wave in free space, since $|\beta| \gg \omega\sqrt{\varepsilon_0 \mu_0}$ in these sub-wavelength structures. As the core slab thickness decreases, this velocity is also reduced and the fields become more concentrated around the interfaces.

In order to provide a more comprehensive understanding of the dispersion in these waveguides, we should take into account the frequency dependence of the ENG material permittivity. We use here the Drude model (including realistic losses) for silver in the IR and visible bands (e.g., [25]), which has a good agreement with experimentally-derived permittivity for silver from 30 to 900 THz (spanning the IR and visible frequencies), in order to evaluate the variation of $\beta$ with the frequency $f = \omega/(2\pi)$ in a realistic dispersive and lossy material. We have assumed $\varepsilon_{Ag} = \left(\varepsilon_\infty - \frac{\omega_p^2}{\omega^2 + i\omega\omega_\tau}\right)\varepsilon_0$ with $\omega_p = 2\pi \cdot 2175\, THz$, $\omega_\tau = 2\pi \cdot 4.35\, THz$ and $\varepsilon_\infty = 5$, following [25]. Fixing the core slab thickness at $d = 14\, nm$ and assuming glass with $\varepsilon_{SiO_2} = 2.19\varepsilon_0$ as the dielectric material (for sake of simplicity we neglect here the dispersion of glass), Fig. 5 shows the $\mathrm{Re}[\beta] - f$ plots for even and odd modes in the two configurations of a $SiO_2$-Ag-$SiO_2$ and an Ag-$SiO_2$-Ag waveguide. The slope of these plots again confirms our prediction regarding the RH and LH behavior of the modes, depending on the value of the real part of the permittivity as a function of frequency (as indicated at selected frequencies).

In Fig, 5a, since we are in the region for which $\mathrm{Re}[\varepsilon_{Ag}] < -\varepsilon_{SiO_2}$, the guided wave number $\beta$ in both cases has a real part that increases with frequency. As $|\mathrm{Re}\,\varepsilon_{Ag}|$ decreases, $\beta$ approaches a vertical asymptote at the frequency where $\varepsilon_{Ag} = -\varepsilon_{SiO_2}$ (if we neglect ohmic losses). This is also seen in Eq. (4) for which $\beta \to \infty$ when $\varepsilon_{in} \to -\varepsilon_{out}$. This is justified by the fact that in the limit where $\varepsilon_{Ag} = -\varepsilon_{SiO_2}$, each of the two interfaces (when isolated) would support a plasmonic surface resonance, and when the mode approaches this limit, its configuration becomes very much concentrated around this interface, effectively creating two isolated resonant lines at $y = \pm d/2$, whose field distribution separates distinctly one resonance from the other. At this limit, the two opposite power flows in the ENG and DPS regions tend to a same absolute value and therefore $|P_{net}| \ll \max(|P_{in}|, |P_{out}|)$, as anticipated in the previous discussion. Most part of the power is in fact reactive in this limit, and the *Q* factor for this resonance is very high. This is also seen in the plots of Fig. 5, since a vertical slope for the curve indicates a zero group velocity, and indeed Eq. (6) yields $v_g = 0$ with $\varepsilon_{ENG} = -\varepsilon_0$. (In the plot here, due to the presence of loss, the slope is never exactly vertical and the point $\text{Re}[\varepsilon_{Ag}] = -\varepsilon_{SiO_2}$ is not a real asymptote. However, since here the losses are moderate, the situation is very similar to the ideal lossless limit as soon as we move away from the resonance peak.)

As the frequency is increased further, we enter Fig. 5b, where backward operation is achieved. In this regime, the $SiO_2$-Ag-$SiO_2$ waveguide operates in its left-handed (i.e., backward) even mode, as predicted in Fig. 1b, whereas the dual geometry has its backward operation with an odd mode (which corresponds to the case analyzed in [15]). We note that for the even case there are two modes, one backward and one forward. When $\varepsilon_{Ag} \simeq 0^-$, Eq. (4) would yield $\beta \simeq 0$, but before reaching that point, the two modes start to interact around $\omega\sqrt{\varepsilon_{SiO_2}\mu_0}$ (the light-line indicated in Fig. 5b). The even mode will no longer be guided inside the ENG core and may lose its guiding properties. This is noticeable in Fig. 5b for the LH NTL, where the slope of the solid line around 880 THz has a drastic change. At this frequency, the forward mode begins to interfere with the backward mode we are interested, and this produces an effect in the dispersion of the backward mode, which is no longer properly guiding the energy, as confirmed in the next Fig. 6b by a sudden increase in its damping factor. Note that around 920 THz, the mode also starts to leak out energy, since the dispersion line crosses the light line, resulting in the increase of the damping factor in Fig. 6b.

In Fig. 6, the corresponding imaginary part of $\beta$, representing the damping factor for these modes, which is due to the material losses in the plasmonic material, is reported. We point out that in this figure the sign of $\text{Im}\,\beta$ is positive for the forward modes and negative for the backward modes, since in Fig. 5 we always considered solutions with positive $\text{Re}\,\beta$ in the $+x$ direction (and therefore for the backward case, one should have a power flow in the opposite direction (i.e., $-x$ direction), implying that in the presence of loss the power flux decays in $-x$ direction, thus "grows" in the $+x$, hence the sign of $\text{Im}\,\beta$). Considering that we have utilized realistic values for material losses and that the cross section of the guided beam is very small (the core thickness is just 14 nm), this example shows that highly confined guided modes can propagate along this structure *without diffraction* for some noticeable distance both in the even and odd mode of operation. It is interesting to note that the losses increase near the asymptote $\varepsilon_{Ag} \simeq -\varepsilon_{SiO_2}$, which is the resonant region, consistent with the fact that in this region the fields are more concentrated (with relatively high values) in a small region near the two interfaces and therefore they are more sensitive to material losses. We do not discuss further the transition into the high-attenuation region beyond 880 THz in the plots for the even mode of operation, since for the purpose of the present paper, the interest is mainly focused on the region of relatively low-damping factors in which NTL modes are present and the circuit analogy may be straightforwardly applied.

It is interesting to note that the NTL circuit analogy, which has been used here for the dominant even mode in these structures, can also be extended to the odd mode of operation. Such odd modes have been utilized by Shvets in [15] for the ENG-air-ENG configuration. In this case the field distribution does not resemble that of a TL with two oppositely directed (antiparallel) wire currents, but instead it looks like that of two nano-wires with similarly directed parallel) currents $I$, which may again be defined in the same manner of Eq. (7). The two modes of operation and the model to describe them are therefore very different, as will be shown in more details in the next section.

As a final remark in this section, it should be pointed out that a chain of plasmonic nano-particles (similar to a single array of nanoparticles we show in middle row of Fig. 1) when excited with an electric field parallel to the chain may, under proper conditions, support a confined guided mode, as has been shown by several groups [26]-[31]. This guided mode has properties similar to the odd mode in the structure shown in Fig. 2. In the circuit analogy, each plasmonic nano-particle would correspond to a lumped nano-inductor, and the vacuum gaps between and around them correspond to the nano-capacitors. Such a cascade of inductors and capacitors is capable of guiding and transmitting the wave energy. An analytical model for such propagation along chains of plasmonic and non-plasmonic particles has been recently derived, and in the limit of closely packed particles corresponds to the predictions of the present analysis. In particular, when the exciting electric field is parallel to the chain axis, the guided mode is consistent with the odd mode, as mentioned above, whereas when the electric field is orthogonal to the chain axis for a single chain or for parallel chains shown in Fig. 1 (top row), the structure deals with the even mode propagation. These issues are not discussed here and some of them will be reported in a future publication.

## Nano-Transmission Lines as 2-D "Flatland" Nano-optics

The geometry studied in the previous section allows the design of NTLs capable of guiding the energy in a structure with sub-wavelength lateral cross section, which, depending on the proper design and desired operation, may be RH or LH. (This may also be achieved by the 1-D cylindrical rods, as found in [17]-[18]) However, as we describe below, the structure analyzed in the previous section may also suggest further interesting and exciting possibilities, particularly for a 2-D optics in the plane of propagation of these anomalous modes.

In the $x-z$ plane of Fig. 2, these guided modes, in fact, are free to propagate with any angle with respect to the $x$ axis. Varying the geometrical parameters of such a waveguide along the $y$ direction affects directly its propagation properties and in particular its phase velocity and its forward or backward operation, i.e., its effective index of refraction in this plane. In other words, for an observer in the $x-z$ plane of propagation any change or modulation of the geometry in the $y$ direction results in a change in the effective refractive index of the planar metamaterial. This is analogous to the concepts suggested in [32], where the terminology "flatland optics" is used, but this relies on a totally different phenomenon, for which effective DPS or DNG planar metamaterials with a desired effective refractive index may be synthesized at optical frequencies, by suitably modifying the geometry of the waveguide along the $y$ axis. This may lead to interesting possibilities for constructing and exploiting materials with anomalous electromagnetic properties at infrared and optical frequencies, as we describe in the following. It is worth underlining an important property of these planar metamaterials: unlike the other common ways of constructing left-handed metamaterials with resonant inclusions, here these effective 2-D planar metamaterials do not rely on a resonant phenomenon, similar to their microwave counterparts synthesized with printed microstrip lines and lumped circuit elements [21]-[22]. This allows the low-loss conditions in their operation, which provides the better possibility for demonstrating some of the unconventional loss-sensitive features of such 2-D metamaterials in the IR and visible frequency domains.

In order to better describe some of the interesting features of these 2-D "flatland" structures, a proper "metric" may be defined to link the geometrical properties of the waveguide in the $y$ direction to its 2-D effective propagation properties in the plane of observation ($x-z$ plane). For an even backward guided mode propagating along the $x$ direction, in the previous section we have already derived the expression (Eq. (7)) for the effective current $I$ (per unit length in $z$) circulating in opposite directions along the two interfaces of such NTL with even mode of operation. By the same token, in analogy with the expression of the current and consistent with the definition of voltage in a classic TL, we may define the effective voltage $V$ (per unit length) at any section of such transmission line as the value of the normal electric field in the $xz$ plane:

$$V = -E_y(y=0). \tag{8}$$

Not surprisingly, this definition satisfies the TL equations for the propagation of voltages and currents in a standard transmission-line: applying Maxwell's equations, in fact, one can find:

$$\frac{dI}{dx} = -\left.\frac{\partial H_z}{\partial x}\right|_{y=0} = -i\omega\varepsilon_{in} \left.E_y\right|_{y=0} = i\omega\varepsilon_{in} V = i\omega\varepsilon_{eff}^{even} V$$
$$\frac{dV}{dx} = -\left.\frac{\partial E_y}{\partial x}\right|_{y=0} = -\left.\frac{\partial E_x}{\partial y}\right|_{y=0} - i\omega\mu_0 \left.H_z\right|_{y=0} = i\omega\mu_{eff}^{even} I \tag{9}$$

with the definitions of:

$$\varepsilon_{eff}^{even} \equiv \varepsilon_{in}$$
$$\mu_{eff}^{even} \equiv \mu_0 + \frac{\left.\partial E_x/\partial y\right|_{y=0}}{i\omega \left.H_z\right|_{y=0}}. \tag{10}$$

In a standard TL, the supported mode is a TEM one, and therefore the effective permittivity and permeability of the propagation are those of the material between the two lines carrying opposite currents. In the NTL case here, however, the propagating even mode is TE, and the mode is distributed in an inhomogeneous region. The effective permittivity remains the same as that of the material inside the line (i.e., $\varepsilon_{in}$ which may be positive or negative), since for this mode the only available component of magnetic field is orthogonal to the propagation direction, but the effective permeability is affected by the presence of the component of the electric field $E_x$ parallel to the propagation vector. As a result, even though all the materials involved here are non-magnetic with permeability $\mu_0$, the *effective* $\mu_{eff}$ may attain a negative value, hence the 2-D structure may behave as a LH structure. From (9) or (10), the dispersion relation for the modes can be expressed as:

$$\beta^2 = \omega^2 \varepsilon_{eff} \mu_{eff}. \tag{11}$$

This confirms again how for a propagating mode (when $\beta^2 > 0$), a positive-permittivity core would imply $\mu_{eff} > 0$ and therefore a positive index of refraction (effectively a RH medium), whereas when $\varepsilon_{in} = \varepsilon_{eff} < 0$, it is required to have $\mu_{eff} < 0$ and therefore a negative effective refractive index for such a planar metamaterial. This is clearly consistent with what was derived in the previous section for the even mode of operation. This analysis, however, allows a relatively easy design of such metamaterial structures, once the effective required parameters are determined.

Continuing with this analogy, we may also define the equivalent characteristic impedance of this NTL, as $Z_c = \sqrt{\mu_{eff}/\varepsilon_{eff}} = \beta/(\omega \varepsilon_{eff})$, which represents the ratio between the equivalent voltage and current at the input port of a matched or infinitely long NTL. It is interesting to note that the quantity $Z_c |I|^2 / 2$, which indicates the power flow into a standard TL, here represents the density of power flowing in the $x-z$ plane, since the following identity holds:

$$Z_c |I|^2 / 2 = \text{Re}[S_x]\big|_{y=0}. \tag{12}$$

We note that this power density in the NTL does not however relate directly to the net power being carried by this structure and actually being available to a user, since there is an oppositely directed power flow in the cladding region. In a standard TL, the characteristic impedance provides information about its matching properties, i.e., two cascaded TL with similar characteristic impedances show a zero reflection coefficient at their junction. Here, however, $Z_c$ represents an effective "averaged" characteristic impedance of the mode and it is not necessarily the same locally point by point along the $y$ axis. This means that two NTL with similar $Z_c$ are not necessarily matched, but may exhibit certain reflection, if not properly designed. Clearly, a necessary condition for a good matching between two NTL is that their modes are similarly distributed in the transverse plane, which is achieved when not only $Z_c$ are similar, but also the transverse thickness of each NTL is comparable. This is why in the design of the structures of Fig. 4 for the RH and LH NTLs, we modeled the lines with the same thickness and similar $Z_c$.

For the odd mode of operation, the analysis should be conducted differently, since the two-wire geometry does not allow a similar effective analogy between voltages and currents along the line. In this case, the parallel currents tend to distribute more widely in the space surrounding the structure, and as a result a valid definition of the effective parameters requires an integration over the space, which may be given following Shvets [15] as:

$$\varepsilon_{eff}^{odd} = \frac{\int_0^\infty \varepsilon E_y dy}{\int_0^\infty E_y dy}$$

$$\mu_{eff}^{odd} = \mu_0 - \frac{E_x\big|_{y=0}}{i\omega \int_0^\infty H_z dy} \tag{13}$$

Also this definition satisfies (11) for the odd modes and is consistent with the right-handed or left-handed properties of these modes as derived in the previous section. (This definition however may not be applied to the even operation.)

For the waveguides of Fig. 4, the effective parameters in the two cases are summarized in Table 1.

From this table, we can see that these structures have high values for *effective* permeability, which is not typically achievable when metamaterials are designed with resonant inclusions. Here, however, these values may be achieved without any active resonance required, and their effective value is high, due to the slow modes they support. Note how the characteristic impedances of the two modes are similar, together with the thicknesses of the two waveguides, implying that a low reflection is expected when the two lines are cascaded (this is also confirmed by the very similar field distribution obtained in Fig. 4). In Fig. 7, we show the effective parameters for the NTLs considered in Fig. 5 and 6 made of silver and glass with material losses considered. We note the low values of the imaginary parts of $\varepsilon_{eff}$ and $\mu_{eff}$, showing how a negative real part of permeability with reasonably low losses may be achieved with this configuration. Furthermore, in Fig. 7b one can see the abrupt change in the effective permeability when the backward mode begins noticing the presence of a higher-order mode, as discussed in the previous section.

It is interesting to point out that the magnitude of the *effective* permeability in such a planar metamaterial is increased when we get close to the resonance of the structure (i.e., for $\varepsilon_{ENG} \simeq -\varepsilon_0$), consistent with the example of Table 1 and the results in Fig. 7. This implies that ohmic losses in the materials eventually limit the maximum magnitude that the effective permeability (positive or negative) can reach, since getting too close to such a resonance may correspond to high damping factors and consequently high imaginary parts for the effective parameters.

Once it is clear that for a "flatland" observer in the $x-z$ plane the structure of Fig. 2 may behave as a DPS or DNG planar metamaterial, we have several possibilities for showing their anomalous wave interaction. The first example may be the one of cascading a DPS material, i.e, a RH NTL as designed in the previous section, with a DNG one, i.e., a LH NTL, in order to verify the negative refraction phenomenon at this interface. We have briefly reported this possibility in [20], utilizing the NTL whose modal field distributions are depicted in Fig. 4 and with geometrical and effective parameters summarized in Table 1. As already noticed, we expect a good matching between the two NTL, owing to their similar characteristic impedances and thickness $d$ of the core slab.

In Fig. 8a, we see the top view of the phase distribution of the electric field on the $x-z$ plane, around the interface between the two NTLs presented in Fig. 4, the left with the RH operation, and the the right with the LH operation. An impinging guided mode, coming from the left, is carrying power obliquely with respect to the interface (at $x=0$). The angle of incidence in this example is selected to be $43°$. A mode-matching technique has been applied, considering the higher-order TM even modes, the radiation modes and the TE odd modes supported by this geometry, which, at the oblique incidence, may couple to the TM even modes under consideration. As clearly seen from this figure, such a mode experiences negative refraction and (almost) total transmission into the LH nano-layer on the right (the reflection coefficient at the interface is around 0.05). In Fig. 8b, a snapshot in time of the electric field distribution is also shown, which clearly shows the strong transmission with negative refraction at this interface. It is interesting to note in the figure the relatively short guided wavelength of this mode due to the slowness of this mode in this structure. We reiterate that this phenomenon, analogous to what has been obtained at microwave frequencies with planar 2-D TL loaded with lumped circuit elements [9, 10], has been obtained here, by using plasmonic isotropic ENG metamaterial layers, with no need for materials with negative permeability, and potentially at frequencies (e.g., IR and visible regimes) where standard lumped circuit elements may not be easily feasible.

In Fig. 9, a plot of the transmission coefficient at the same interface, as a function of $\beta_z / \omega\sqrt{\varepsilon_0\mu_0}$ is shown, where $\beta_z$ is the transverse component of $\beta$ (transverse with respect to the interface at $x=0$). Here the values of $\beta_z$ for which a very high transmission is possible at this interface is much higher than $\omega\sqrt{\varepsilon_0\mu_0}$; however, we must remember that for our design the value of $\beta$ for the guided modes in the two lines is already much higher than $\omega\sqrt{\varepsilon_0\mu_0}$: $\beta_{RH} = 29.5 k_0$, $\beta_{LH} = 28.95 k_0$. Nevertheless, the plot of transmission coefficient shows the realistic possibility of sub-wavelength focusing with this setup, since when $\beta_z$ becomes greater than $\beta_{LH}$, and the corresponding wave becomes evanescent in the $x$ direction, the "growing exponential" phenomenon [1]-[2] may be achieved in these NTLs, as evident form the growth of the transmission coefficient in this evanescent region in Fig. 9.

The potential applications of this structure may span from sub-wavelength focusing and near-field imaging to diffraction-less guided transport of signal. In order to numerically verify the possibility of sub-wavelength focusing in this planar set-up, in our theoretical study we place a source at a given distance from such an interface, expecting to obtain a focusing on the other side of the interface, with a resolution much higher than that in free space, mainly due to the negative refraction of the propagating modes (which already spam a range of spatial frequencies much higher than that in vacuum, as already noticed from Fig. 8), and also due to the growing-exponential phenomenon previously predicted, which may allow, in principle a higher-than-conventional resolution, similar to what was discussed in [2]. In Fig. 10, a localized current (i.e., a very small source in the observation plane) has been placed at the location $(x_s = -0.15\lambda_0, z_s = 0)$ and the phase distribution and a snapshot in time of the instantaneous electric field amplitude has been shown in the figure. An image of the source is reconstructed on the other side of the interface, and the focusing properties of such an interface are clearly seen, as predicted. Again, the resolution is much higher than the one achievable in free-space optics.

In Fig. 11, we compare the source distribution at the point $x = -0.15\lambda_0$ in the RH NTL with the image distribution at $x = 0.15\lambda_0$ for different cases: (a) when the half-space $x > 0$ is filled with the LH NTL (corresponding to the solid line), as in the case of Fig. 10; (b) when the entire space is filled with the RH NTL (and therefore there is no focusing); (c) when everything would be in vacuum; (d) when we have just the RH NTL with an ideal optical lens to focus the source plane into the image plane; and finally (e) when we have the same lens but in free space. As you can see the RH-LH interface restores the source plane into the image plane with a very good resolution, which is much higher than that in the free-space. As mentioned earlier, this is mainly due to the negative refraction and focusing of the propagating modes.

These RH and LH NTLs can be extended to more complex geometries such as the one in Fig. 12, which is indeed the planar plasmonic version of Pendry's lens [2], consisting of an LH NTL segment sandwiched between two RH NTLs. This can provide a planar imaging system in the "flatland" optics of the x-z plane with a resolution higher than the conventional values. Other interesting possibilities to envision include structures such as small segments of paired RH-LH NTLs in the 2-D planar structures with high scattering cross sections. This will be analogous to the high scattering scenarios we studied in DPS-DNG concentric spheres and cylinders in [7].

## Conclusions

Motivated by our development of nano-scale circuit elements at the optical frequencies, here we have suggested the concept of optical nano-transmission lines and we have shown how to synthesize such NTLs by stacking planar slabs of plasmonic and non-plasmonic materials, which can, under proper designs, exhibit the properties of forward (right-handed) or backward (left-handed) wave propagation. Negative refraction and left-handed propagation can be manifested in such structures without any need for materials with negative permeability. The guided wave propagation in these NTLs have been studied in detail, effective material parameters for the guided modes in such structures have been derived, and some of the potential applications such as sub-wavelength focusing and imaging in these planar optics have also been analyzed.

## Acknowledgements

This work is supported in part by the U.S. Air Force Office of Scientific Research (AFOSR) grant number FA9550-05-1-0442. Andrea Alù has been partially supported by the 2004 SUMMA Graduate Fellowship in Advanced Electromagnetics.

## Author Contact Information

Nader Engheta is the corresponding author (engheta@ee.upenn.edu, phone: +1-215- 898-9777, fax: +1-215-573-2068).

*Accepted for publication in JOSA B Focus Issue on Metamaterials, to appear in March 2006 Issue*[9] R. A. Shelby, D. R. Smith, and S. Schultz, "Experimental verification of a negative index of refraction," *Science* **292**, 77-79 (2001).

[10] L. Landau, and E. M. Lifschitz, *Electrodynamics of continuous media* (Elsevier, 1984).

[11] V. A. Podolskiy, A. K. Sarychev, and V. M. Shalaev, "Plasmon modes in metal nanowires and left-handed materials," *Journal of Nonlinear Optical Physics and Materials* **11**, 65-74 (2002).

[12] S. O'Brien, D. McPeake, S. A. Ramakrishna, and J. B. Pendry, "Near-infrared photonic bandgaps and nonlinear effects in negative magnetic metamaterials," *Physical Review B* **69**, 241101 (2004).

[13] G. Shvets, and Y. A. Urzhumov, "Engineering electromagnetic properties of periodic nanostructures using electrostatic resonances," *Physical Review Letters* **93**, 243902 (2004).

[14] M. L. Povinelli, S. G. Johnson, J. D. Joannopoulos, and J. B. Pendry, "Toward photonic-crystal metamaterials: creating magnetic emitters in photonic crystals," *Applied Physics Letters* **82**, 1069-1071 (2003).

[15] G. Shvets, "Photonic approach to making a material with a negative index of refraction," *Physical Review B* **67**, 035109 (2003).

[16] A. Alù, A. Salandrino, and N. Engheta, "Negative effective permeability and left-handed materials at optical frequencies," submitted for publication, under review. (The manuscript can be viewed online at: http://arxiv.org/pdf/cond-mat/0412263.)

[17] A. Alù, and N. Engheta, "Anomalies in the surface wave propagation along double-negative and single-negative cylindrical shells," presented at the *2004 Progress in Electromagnetics Research Symposium (PIERS'04)*, Pisa, Italy, March 28-31, 2004, CD Digest.

[18] J. Takahara, S. Yamagishi, H. Taki, A. Morimoto, and T. Kobayashi, "Guiding of a one-dimensional optical beam with nanometer diameter," *Optics Letters* **22**, 475-477 (1997).

[19] N. Engheta, A. Salandrino, and A. Alù, "Circuit elements at optical frequencies: nano-inductors, nano-capacitors and nano-resistors," *Physical Review Letters* **95**, 095504 (2005).

[20] A. Alù, and N. Engheta, "Sub-wavelength focusing and negative refraction along positive-index and negative-index plasmonic nano-transmission lines and nano-layers," *Proceedings of the 2005 IEEE Antennas and Propagation Society (AP-S) International Symposium*, Washington, DC, USA, Vol. 1A, pp. 35-38, July 3-8, 2005.

[21] G. V. Eleftheriades, A. K. Iyer, and P. C. Kremer, "Planar negative refractive index media using periodically L–C loaded transmission lines," *IEEE Transactions on Microwave Theory and Techniques* **50**, 2702-2712 (2002).

[22] L. Liu, C. Caloz, C.-C. Chang, and T. Itoh, "Forward coupling phenomena between artificial left-handed transmission lines," *Journal of Applied Physics* **92**, 5560-5565 (2002).

[23] A. Alù, and N. Engheta, "Mode excitation by a line source in a parallel-plate waveguide filled with a pair of parallel double-negative and double-positive slabs," *Proceedings of 2003 IEEE Antennas and Propagation Society (AP-S) International Symposium*, Columbus, OH, USA, Vol. 3, pp. 359-362, June 22-27, 2003.

[24] Note that the current distribution in a transmission line as in Figs. 1a and 1b, would indeed excite a magnetic field distribution with even distribution with respect to the transverse coordinate, thereby justifying the use of the even mode in the ENG-DPS-ENG waveguide (for RH behavior) and in the DPS-ENG-DPS waveguide (for LH behavior)

[25] I. El-Kady, M. M. Sigalas, R. Biswas, K. M. Ho, and C. M. Soukoulis, "Metallic photonic crystals at optical wavelengths," *Physical Review B* **62**, 15299-15302 (2000).

[26] M. Quinten, A. Leitner, J. R. Krenn, and F. R. Aussenegg, "Electromagnetic energy transport via linear chains of silver nanoparticles," *Optics Letters* **23**, 1331-1333 (1998).

[27] S. A. Tretyakov, and A. J. Vitanen, "Line of periodically arranged passive dipole scatterers," *Electrical Engineering* **82**, 353-361 (2000).

[28] S. A. Maier, M. L. Brongersma, and H. A. Atwater, "Electromagnetic energy transport along arrays of closely spaced metal rods as an analogue to plasmonic devices," *Applied Physics Letters*, **78**, 16-18 (2001).

[29] A. D. Yaghjian, "Scattering-matrix analysis of linear periodic arrays," *IEEE Transactions on Antennas and Propagation* **50**, 1050-1064 (2002).
-16-

**Tables**

Table 1 – Effective parameters for the NTL of Fig. 4

| Parameter | RH NTL (left in Fig. 4) | LH NTL (right in Fig. 4) |
|---|---|---|
| $\varepsilon_{in}$ | $\varepsilon_0$ | $-0.95\,\varepsilon_0$ |
| $\varepsilon_{out}$ | $-1.05\,\varepsilon_0$ | $\varepsilon_0$ |
| $d$ | $\lambda_0/50$ | $\lambda_0/50$ |
| $\varepsilon_{eff}$ | $\varepsilon_0$ | $-0.95\,\varepsilon_0$ |
| $\mu_{eff}$ | $870.432\,\mu_0$ | $-882.763\,\mu_0$ |
| $\beta$ | $29.5\,k_0$ | $28.95\,k_0$ |
| $Z_c$ | $29.5\,Z_0$ | $30.5\,Z_0$ |

**Figure Captions**

Fig. 1. (color online) Conceptual synthesis of right-handed and left-handed nano-transmission lines (NTL) at optical frequencies. (Top row): Conventional circuit model of RH and LH lines using distributed inductors and capacitors; (middle row): Plasmonic and non-plasmonic nano-particles may play the role of nano-inductors and nano-capacitors, following [18]; (bottom row): Closely packed nano-particles, in the limit, become plasmonic and dielectric layers, which may be employed in a similar way at a NTL. A sketch of the voltage ($V$) and current ($I$) symbols along the lines is also depicted.

Fig. 2. Geometry of the 2-D nano-transmission line (NTL) analyzed here: a core slab with permittivity $\varepsilon_{in}$ and thickness $d$ sandwiched between two half-spaces with permittivity $\varepsilon_{out}$.

Fig. 3. Dispersion properties of the odd and even guided modes supported by the structure of Fig. 2 with: a) $\varepsilon_{in} = -5\varepsilon_0$, $\varepsilon_{out} = \varepsilon_0$ (odd mode) and reversing the two materials (even mode); b) $\varepsilon_{in} = \varepsilon_0$, $\varepsilon_{out} = -\varepsilon_0/5$ (odd mode) and reversing the two materials (even mode), following the exact Eq. (3) and the approximate Eq. (4) valid for electrically thin layers, and assuming no material loss. The guided wave number $\beta$ increases hyperbolically when the thickness decreases, leading to a concentration of the guided mode around the two interfaces for sub-wavelength slabs. In the figure axes, $k_0 = \omega\sqrt{\varepsilon_0\mu_0}$ is the free-space wave number and $\lambda_0 = 2\pi/k_0$ is the corresponding free space wavelength.

Fig. 4. Field distributions for the dominant TM even modes in the two cases of RH and LH NTL (cross section is shown here). The parameters have been chosen to have the same core thickness $d = \lambda_0/50$) and similar field distributions, and in the two cases $\beta_{RH} = 29.5\,\omega\sqrt{\varepsilon_0\mu_0}$, $\beta_{LH} = 28.95\,\omega\sqrt{\varepsilon_0\mu_0}$. The $\mathbf{P}_{net}$ vector indicates the direction of net power propagation, whereas the $\beta$ vector refers to the phase flow.

Fig. 5. Dispersion plots for the geometry envisioned in Fig. 2 taking into account the Drude model for silver (including loss): a) the positive slope confirms a forward behavior for the RH NTL (even mode for the structure of Fig. 1a) and for the dual geometry and the odd mode; b) here the negative slope indicates the LH (i.e., backward) operation of these structures. The variation of permittivity for the ENG material is indicated in the Figure at selected frequencies.

Fig. 6. Damping factors ($\text{Im}\,\beta$) for the dispersion plots of Fig. 5, due to the presence of ohmic losses. Note how the damping factor has different signs in the two cases, due to the backwardness of the structures in b). Real and imaginary parts of the silver permittivity are reported in the Figure at certain selected frequencies.

Fig. 7. The effective material parameters for the NTL considered in Figs. 5 and 6: a)ENG-air-ENG waveguide as a RH NTL; b) air-ENG-air waveguide as a LH NTL. Note how an *effectively* positive or negative permeability is achieved, even when realistic material losses are assumed (as in Figs. 5 and 6) and the materials are all non-magnetic.

Fig. 8. (color online) Interface between the RH and LH NTLs of Fig. 4 (a) top view of the contour plot of the phase distribution; b) 3-D plot of the instantaneous electric field distribution at the *x-z* plane). Note the nearly complete transmission, and the low reflection and clearly evident negative refraction at the interface, which underlines the negative refraction (i.e., "left-handedness") of the second NTL and the good matching between the two structures.

Fig. 9. Transmission coefficient at the interface between the RH and LH NTLs of Fig. 4, as a function of the transverse $\beta_z$ (where $\sin^{-1}(\beta_z/\beta)$ represents the angle of incidence of the impinging mode). Note the growing transmission coefficient in the evanescent region, connected to the *growing-exponential* and sub-wavelength focusing phenomenon (e.g., [1]-[2]).

Fig. 10. (color online) The same as in Fig. 8, but for a small localized source in the RH NTL. The sub-wavelength focusing and image reconstruction on the other side of the interface are clearly noticed.

Fig. 11. (color online) Analysis of imaging features due to the RH-LH interface described in the previous figures. Sub-wavelength resolution is clearly noticeable in this analysis. More details are given in the text.

Fig. 12. A sketch of a proposed planar set-up as a sub-wavelength imaging system, which consists of an LH nano-transmission line sandwiched between two RH NTLs, for operation in the IR and visible domains.

**Figures**

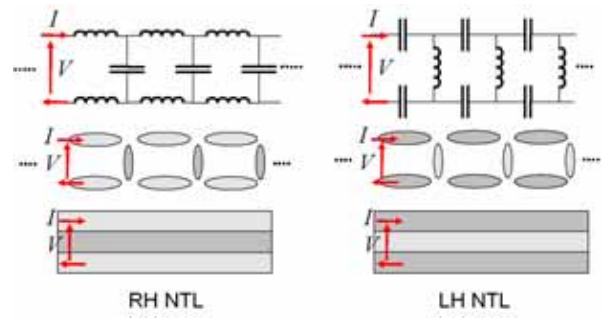

Figure 1

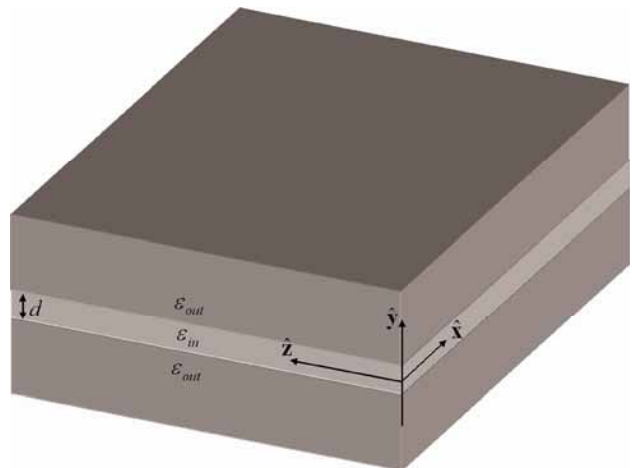

Figure 2

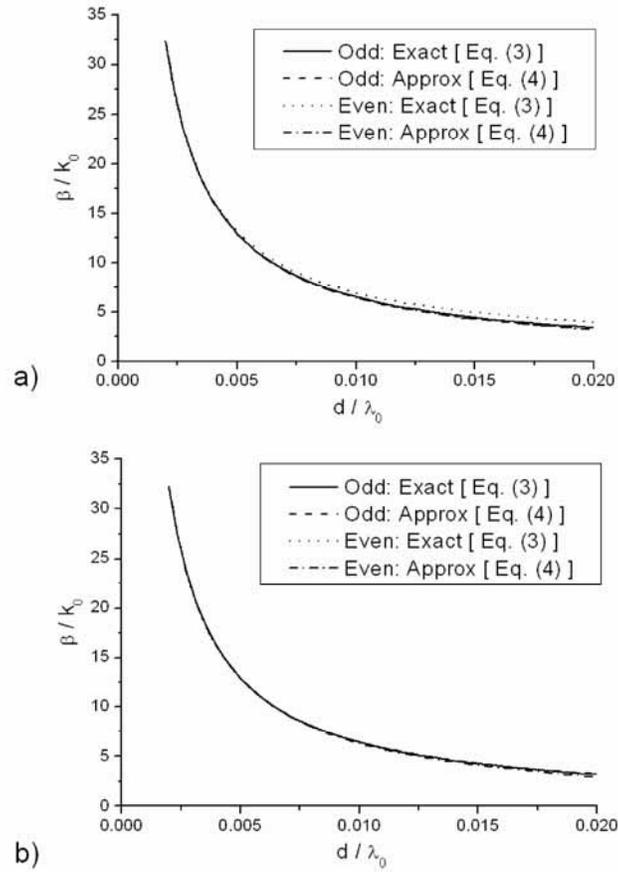

Figure 3

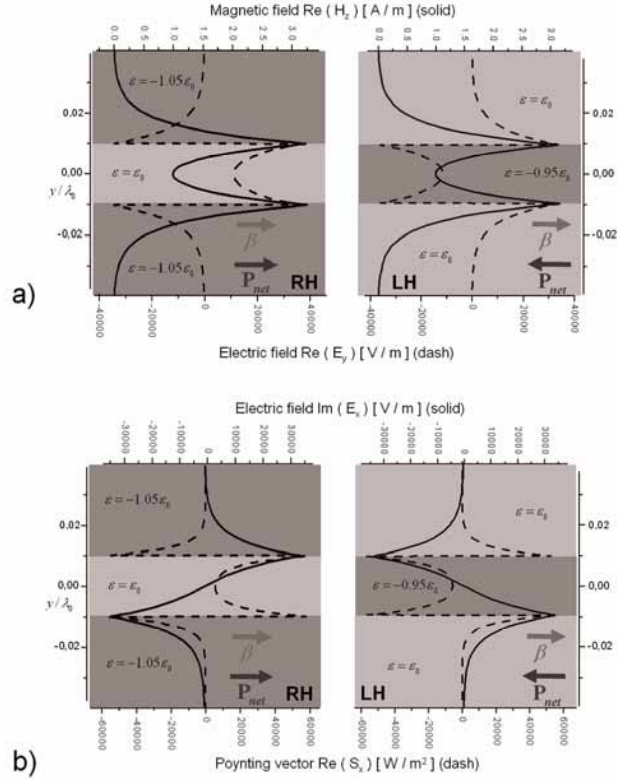

Figure 4

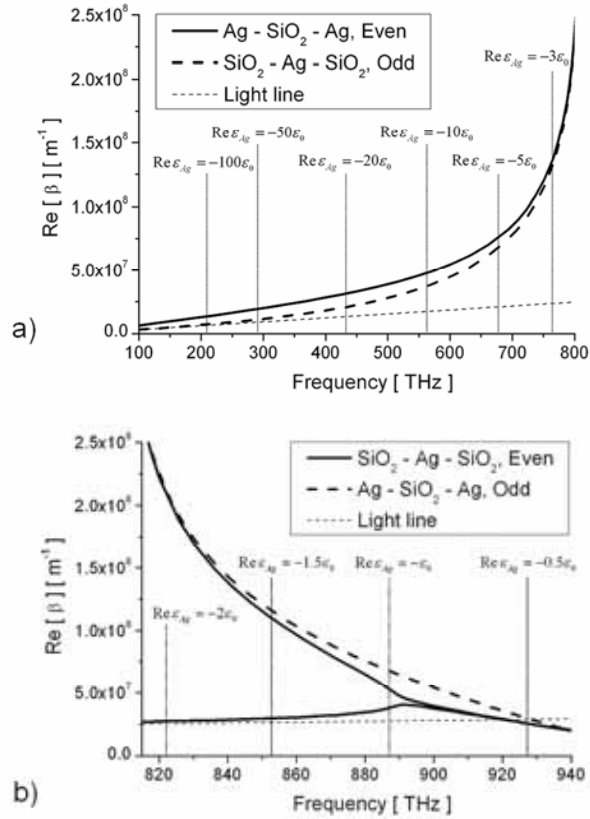

Figure 5

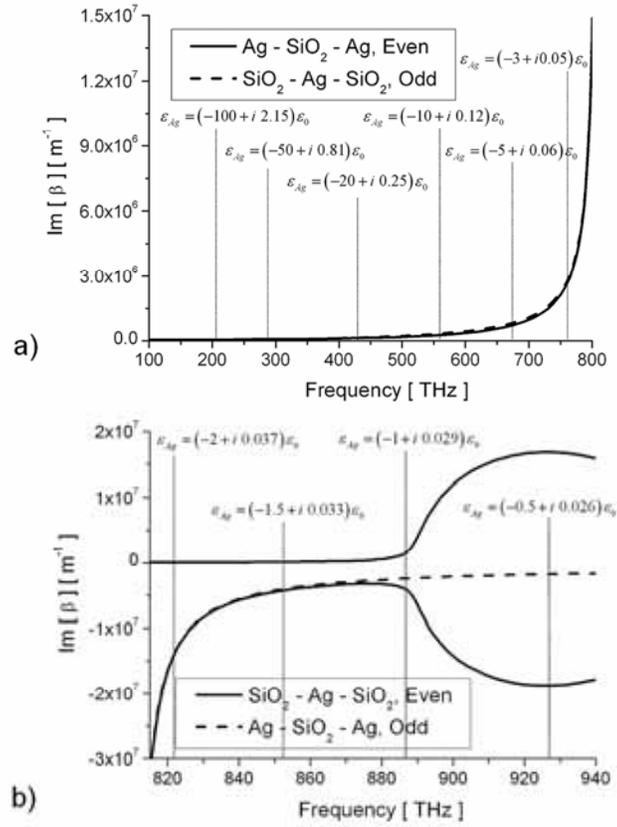

Figure 6

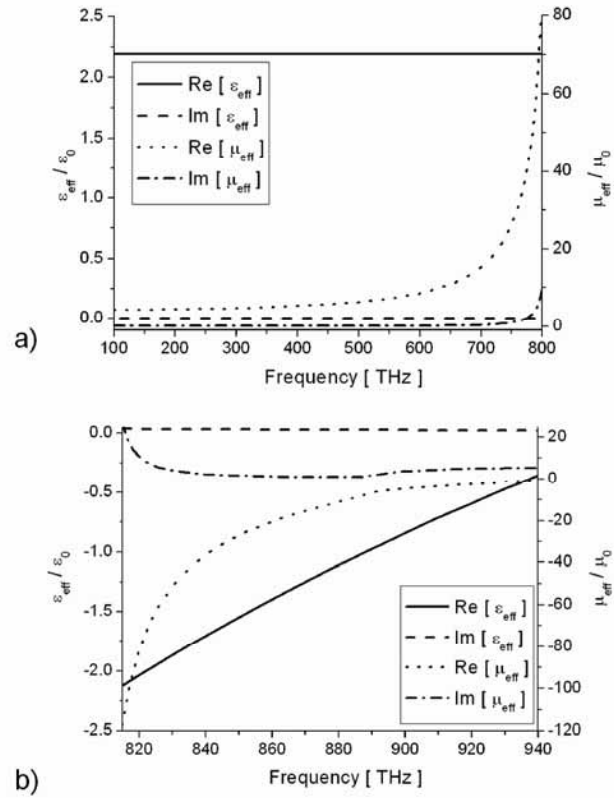

Figure 7

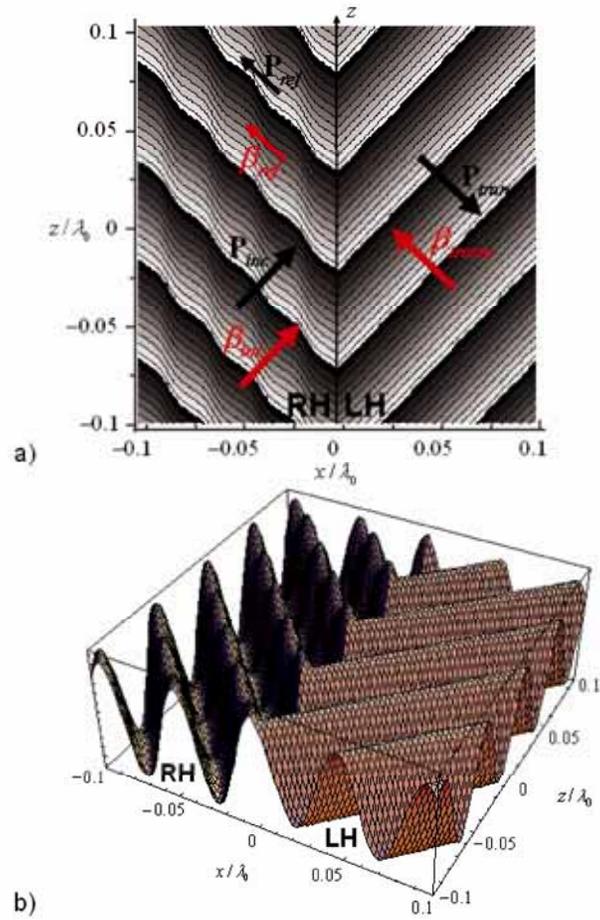

Figure 8

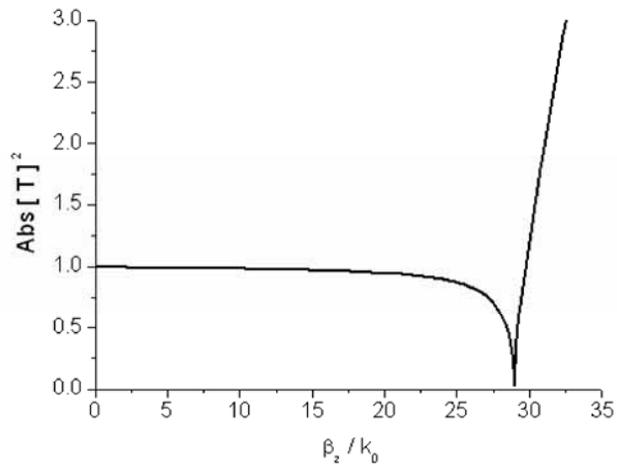

Figure 9

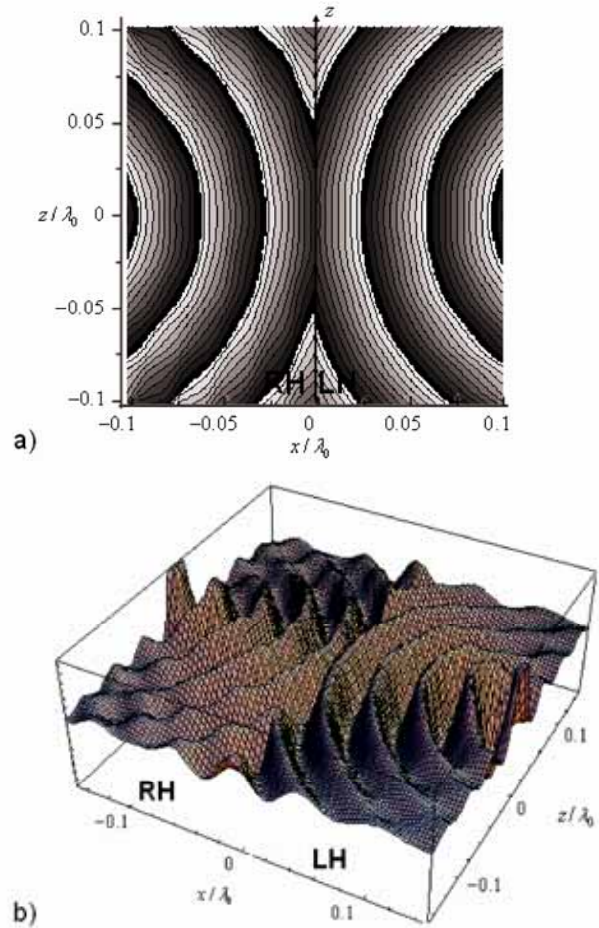

Figure 10

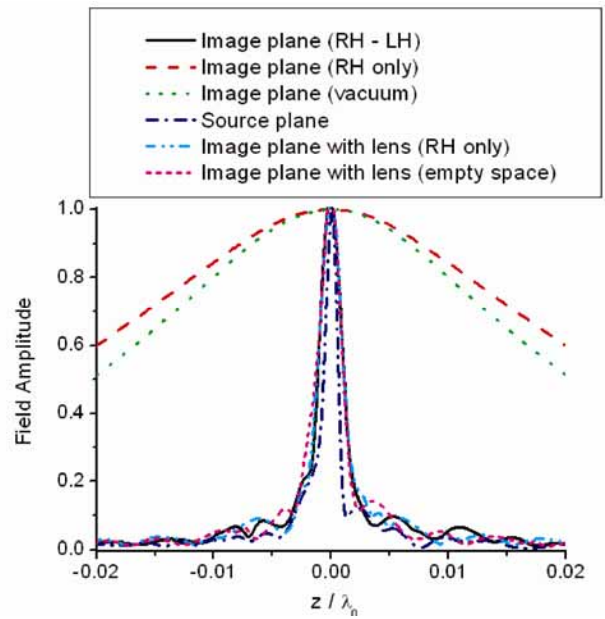

Figure 11

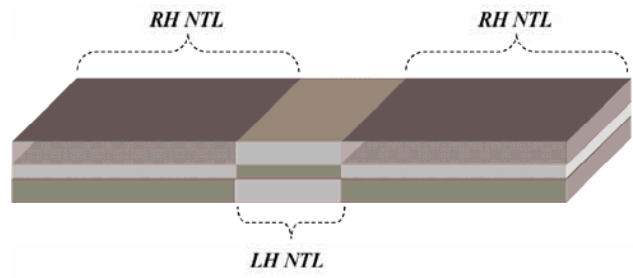

Figure 12